\documentstyle[prd,aps]{revtex}

\begin{document}
\draft
\twocolumn[\hsize\textwidth\columnwidth\hsize\csname
@twocolumnfalse\endcsname
\preprint{PACS: 98.80.Cq, DTP/97/92, IJS-TP-97/16, 
OUTP-97-49-P,\\
{}~hep-ph/yymmnn}

\newcommand\lsim{\mathrel{\rlap{\lower4pt\hbox{\hskip1pt$\sim$}}
    \raise1pt\hbox{$<$}}}
\newcommand\gsim{\mathrel{\rlap{\lower4pt\hbox{\hskip1pt$\sim$}}
    \raise1pt\hbox{$>$}}}

\title{Large Lepton Number of the Universe and 
The Fate of Topological Defects}

\author{Borut Bajc$^{(1,3)}$,  Antonio
Riotto$^{(2,3)}$ and Goran Senjanovi\'c$^{(3)}$}

\address{$^{(1)}${\it Department of Physics, University of Durham, 
Durham DH1 3LE, UK}, and 
{\it J. Stefan Institute, 1001 Ljubljana, Slovenia}}

\address{$^{(2)}${\it Department of Physics, Theoretical Physics, 
University of Oxford, 1 Keble Road, Oxford OX1 3NP, UK}}

\address{$^{(3)}${\it International Center for Theoretical Physics,
34100 Trieste, Italy }}

\date{\today}
\maketitle

\begin{abstract}
We show that the monopole problem in Grand Unified Theories as 
well as the domain wall problem may be easily solved if the lepton 
number asymmetry in the Universe is 
large enough.
\end{abstract}

\pacs{PACS: 98.80.Cq    \hskip 1 cm DTP/97/92, IJS-TP-97/16, 
OUTP-97-49-P}
\vskip1pc]

{\it  A. Introduction}. \hspace{0.5cm}  Monopole and domain wall problems 
are some of the central issues in the modern astroparticle physics. The
problem of monopoles is especially serious since it is generic to the idea
of grand unification \cite{p79}. The popular solution based on the
idea of inflation  
cannot be implemented in the minimal Grand Unified Theories (GUTs), and even 
if it does work it 
would imply a sad prediction of essentially no monopoles in the Universe,
and thus eliminate a prospect of observing this exciting aspect of charge
quantization. 
Of course, it is hard to imagine a Universe without ever
having passed through an era of inflation; we simply take here the point
of view that this may have happened before the time of grand unification
in the thermal history of the Universe. Similarly, the problem of domain
walls \cite{zko74} in theories with a spontaneous breaking of 
discrete symmetries
requires inflation to take place after the   phase
transitions that cause the production of these defects, which is
difficult to achieve in general. 
Recently, a possible solution of the monopole 
problem was suggested \cite{gv97}, based on the possibility that 
unstable domain walls sweep away the monopoles. 

There is another possible way out of these problems  and it is based 
on an unusual picture of non restoration of symmetries at high temperatures. 
It has been known for a long time that in theories with more than 
one Higgs multiplet, which seems to be a necessary feature of all theories 
beyond the Standard Model (SM), broken symmetries 
may remain broken at high temperature $T$ 
in some regions of the parameter space, and even the unbroken ones
may get broken as the system in question is heated up
\cite{w74,ms79}. 

The idea of symmetry nonrestoration  provides a simple way out of the 
domain wall problem \cite{ds95,dms96}, but unfortunately in the case 
of the monopole problem the situation is far from clear \cite{dms95}, 
since next to the leading order effects tend to invalidate 
this picture for local symmetries \cite{bl95b}. While 
in the case of discrete symmetries the original lattice 
calculations spoke against nonrestoration at high $T$ \cite{bitu97}, 
the latest results give full support of this idea \cite{jl98}, 
as do the other nonperturbative results \cite{roos96}. 
However, it can be shown that this scenario does not work in
supersymmetry. More precisely, there is a rigorous proof at the
renormalizable level \cite{h82,m84}, and the simple counter examples at the
non-renormalizable level \cite{dt96} have been shown not to work \cite{bms96}.

A manifestation of nonrestoration is an old idea \cite{lp80} of 
$U(1)_{em}$ breaking at temperatures above $M_W$. Unfortunately, this 
suffers from the same next-to-leading order effects mentioned above 
\cite{dms96}. There is a simple variation of this scenario where 
$U(1)_{em}$ is broken only in a very narrow range of temperatures 
around the electroweak scale \cite{ds92}, \cite{fkwy92}. In this 
case the monopoles get produced with the hope of being annihilated 
fast enough through the strings attached to monopole-antimonopole 
pairs. However, there is a serious question whether the annihilation 
does the job \cite{gkt92}, \cite{hkr92}.

The situation becomes much more promising if one accepts the
possibility of having a large background charge in the Universe, 
large in a sense of being comparable to the entropy \cite{l76}. 
The presence of some 
sizeable charge asymmetry may postpone symmetry restoration 
in nonsupersymmetric theories  \cite{hw82} or, even more remarkably, 
it can lead to symmetry breaking of internal symmetries at high 
temperature \cite{bbd91}. Furthermore, the phenomenon 
of symmetry nonrestoration at high $T$ in presence of 
large charge asymmetries  has been 
recently shown to work in supersymmetry too \cite{rs97}. 
The principal candidate for a large charge is the  lepton number which 
today could reside in the form of 
neutrinos. This has inspired Linde in his original work to point out that 
large enough lepton number of the Universe would imply the nonrestoration 
of symmetry even in the SM \cite{l79}. While one could 
naively think that the large lepton number 
would be washed out by the sphaleron effects
at the temperature above the weak scale, it turns out  that the
nonrestoration of symmetry prevents this from happening \cite{ls94},
and remarkably 
enough up to this day this still remains a consistent possibility. 
Indeed, the successful predictions of primordial nucleosynthesis are 
not jeopardized as long as the lepton number is smaller than 
$\sim 7\: T^3$ at temperatures of the order of 1 MeV \cite{ks92}. 
It is therefore natural 
to ask ourselves whether a large lepton asymmetry in the 
Universe may play any significant role in solving the monopole problem. 
The main point we wish to make in this Letter  is that the answer may 
be  positive if these two basic requirements are satisfied: the large
lepton asymmetry leads to  
the symmetry nonrestoration of the SM gauge symmetry and  some 
charge field condensation takes place. While it is not clear 
whether this happens in SM, it is certainly true for its 
minimal extensions (such as an additional charged scalar). 
Thus, if the lepton number of the Universe were to turn 
out large, there would be no monopole problem whatsoever.

Now, if Nature has chosen the option that the lepton number 
is large enough so that SM symmetry is not restored at high $T$, 
but without any charge field 
condensation,  even in this case the cosmological consequence 
would be remarkable, for this would suffice to nonrestore the symmetry in 
the minimal model of spontaneous CP violation with two Higgs doublets 
\cite{l73}.
Namely, without the
external charge, in this particular model CP
is necessarily restored at high temperature \cite{dms96} 
leading to the domain wall problem.

As we mentioned before, it was shown recently that 
the phenomenon of nonrestoration at high $T$ in the 
presence of a large charge works in supersymmetry too. 
We have exemplified our findings on simple U(1) models 
\cite{rs97}. It can be shown that this is true in general, 
and all that we say above works also in Minimal Supersymmetric 
Standard Model (MSSM). In this Letter we wish to avoid any 
model building, but rather concentrate on SM showing that 
its cosmology may be something entirely different from what 
one normally imagines.

\vspace{0.2cm}
{\it B. Large L and high T symmetry nonrestoration}\hspace{0.5cm}
Let us now discuss in some detail what happens at high temperature if
the lepton number is large. Notice first that, since we can assume that
the lepton number $L$ is conserved (the sphaleron effects are 
suppressed \cite{ls94}), then the ratio of the lepton density 
$n_L$ to entropy density $s$ is constant, too. Now, we are interested 
in the temperatures above the weak scale when the number of light degrees 
of freedom grows by another order of magnitude with respect to 
$T\simeq 1$ MeV. Thus, the above cited limit $n_L/T^3< 7$ at the 
time of nucleosynthesis becomes for us an order of magnitude bigger: 
$n_L/T^3<70$.

In order to study symmetry breaking, we need to compute the 
effective potential at high $T$ and high chemical potential. 
We employ the approximation $\mu_L < T$ (where $\mu_L$ is the 
chemical potential associated with the lepton number), since 
in this case one can obtain the solutions in a closed form. 
With increased $\mu_L$ the physical effect of symmetry 
breaking gets only stronger \cite{l79}.

The baryon number of the Universe is negligible, $n_B/s \simeq 
10^{-11}$, thus we work in the approximation  $B=0$. Since the 
gauge potentials act essentially as the  chemical potentials 
at high $T$, we include them in our $V_{eff}(T,\mu)$. 
A word of caution is in order. Although $B=0$, since the quarks carry 
non trivial baryon number, one must include the associated chemical 
potential $\mu_B$ and we will see below that it does not vanish. We 
will see that quarks carry a nonvanishing electric charge at high $T$, 
similarly to the $W$ bosons and the charged Higgs scalar.

Using the techniques of \cite{hw82,bbd91} the effective potential 
at high  $T$ ($T >\mu \gg M_W$) and large $n_L$ for the Higgs doublet 
$H$ in the direction of its neutral vacuum expectation value 
$H=(0, v/\sqrt{2})^T$ reads

\begin{eqnarray}
\label{veffall}
V_{eff}&=&\lambda{v^4\over 4}+
{g^2\over 2}\left[(A_0^aA_0^b)(A_i^aA_i^b)-(A_0^aA_0^a)
(A_i^b A_i^b)\right]\nonumber\\
&+&{g^2\over 4}\left[(A_i^aA_i^a)(A_j^bA_j^b)-(A_i^aA_i^b)
(A_j^aA_j^b)\right]\nonumber\\
&-&{v^2\over 8}\left[g^2(A_0^aA_0^a)+g'^2
(B_0B_0)\right]+{gg'\over 2}B_0A_0^3{v^2\over 2}\nonumber\\
&+&{g^2\over 4}\left[A_i^1A_i^1+A_i^2A_i^2\right]{v^2\over 2}
+\mu_Ln_L\nonumber\\
&-&T^2\left({\mu_L^2\over 4}+{\mu_B^2\over 9}\right)+
{T^2\over 3}(g'B_0)\left(\mu_L-{\mu_B\over 3}\right)\nonumber\\
&-&{13\over 36}(g'B_0)^2T^2
-{7\over 12}g^2A_0^aA_0^aT^2+\lambda'T^2{v^2\over 2}.
\end{eqnarray}
where  $A_\mu^a$ and $B_\mu$ are the $SU(2)_L$ and $U(1)_Y$ gauge 
potentials. In the above we have used $g' B_i = g A_i^3$ which 
follows from the equation of motion for $B_i$ and

\begin{equation}
\lambda'={1\over 12} \left[6\lambda+
y_e^2+3y_u^2+3y_d^2+
{3\over 4}(g'^2+3g^2)\right], 
\end{equation}

\noindent
where $y_f$ are the fermionic Yukawa couplings. For simplicity, 
we take only the third generation of fermions since its couplings 
are dominant. The inclusion of the first two generations is 
straightforward and does not change our conclusions.

Notice the point we made before. Although we take $B=0$ , the
associated chemical potential plays an important role in the above
expression. The equations of motion  for the gauge fields $A_\mu^a$ 
show that the solution discussed in \cite{l79} -- all 
gauge potentials zero except for $A_0^3$ and $A_1^1$ -- is consistent with 
the above constraints. 

Using the constraints $\partial V_{eff}/\partial x=0$ for 
$x=\mu_L$, $\mu_B$, $g'B_0$ and $gA_0^3$ we can rewrite the 
effective potential as a function of $v$ and $C=\langle A_1^1\rangle$
only:

\begin{eqnarray}
V_{eff}&=&{\lambda\over 4}v^4 +{\lambda'\over 2}  T^2 v^2 +
{g^2\over 8}v^2 C^2+{n_L^2\over T^2}\nonumber\\
&+&{4 n_L^2(3 v^2 + 12 C^2 + 14 T^2) 
\over 54 v^2 C^2 + (87 v^2 + 96 C^2) T^2 + 112 T^4}\;.
\end{eqnarray}
The effective potential is manifestly bounded from below and it is a
simple exercise to minimize it. We work with a small
$(H^\dagger H)^2$ coupling -- only for the sake of presenting simple analytic
expressions.  We find the results
presented below.
In discussing them, it will turn out useful to  have the individual
distribution of the various charges. Namely, here lies an important
point that was overlooked before \cite{l79} and that plays a significant role
for our considerations about the monopole problem, {\it i.e.} the fact that 
quarks,  the charged Higgs and $W$ carry electromagnetic charge in spite
of having lepton number zero. 
Since the $\mu$-dependent part 
of the effective potential can be written \cite{hw82,bbd91}
\begin{equation}
\label{veffmu}
V_{eff}^{(\mu)}=-{T^2\over 12}\sum_{f}\mu_i^2-{T^2\over 6}
\sum_{b}\mu_i^2-\sum_{b}\mu_i^2|\phi_i|^2+\mu_Ln_L,
\end{equation}
one can find for the distribution of fermionic and bosonic charges

\begin{eqnarray}
\label{qf}
({\cal Q}_F^a)_i&=&q_i^a\mu_i\left({T^2\over 6}\right)\;,\\
\label{qb}
({\cal Q}_B^a)_i&=&q_i^a\mu_i\left({T^2\over 3}+2|\phi_i|^2\right),
\end{eqnarray}
where $q_i^a$ denote the transformation property $
\varphi_i \rightarrow e^{i q_i^a T^a } \varphi_i$,  $\varphi_i$ stands 
for any field, $a$ goes over all the relevant
charges ($L,B, Y/2, T_{3W}$) and $\mu_i=\sum_aq_i^a\mu_a$.

Let us first shortly discuss the case of small lepton asymmetry or, 
more precisely,  $n_L < (n_L)_1 \equiv (4/3) \sqrt{\lambda'} T^3$. In
 such a case 
only the trivial solution is possible: $v = C = 0$. 
This is the usual scenario of small charge densities. 

It is an easy exercise to compute the distribution of charges. One
finds $
L (\nu_L) = L (e_L) = {3 \over 8} L$ and $ L (e_R) = {2 \over 8} L$
for the $L$ number distribution (notice that $L({\nu_L}) = L({e_L})$
since $SU(2)$ is not broken). For the electromagnetic charge (we
list only the nonvanishing ones)

\begin{eqnarray}
Q(e_L) = -{3 \over 8} L & \;,\; Q(e_R) = -{2 \over 8} L & \;,\nonumber \\
Q(u_R) = {2 \over 8} L & \;,\; Q(d_R) = {1 \over 8} L & \;,\; 
Q(h^+) = {2 \over 8} L \;,
\end{eqnarray} 
so that $Q_{tot} =0$ as it should be, but the charge is distributed
among both  fermions and charged Higgs bosons (we find later $W^{\pm}$
participating too).

Let us now focus on the following intermediate range 
$(n_L)_1 < n_L < (n_L)_2 \equiv 
(n_L)_1( 1 + {203 \over 192}g^2 /\lambda')$. It is easy to show that now 
$v \neq 0$, but $C$ still vanishes

\begin{eqnarray}
\label{vb}
V_{eff}(v\neq 0)&=&V_{eff}(v=0)-{21\over 58T^2}
[n_L - (n_L)_1]^2\;,\nonumber\\
v^2&=&{112\over 87}{n_L-(n_L)_1\over 
(n_L)_1}T^2,\;\;\; C=0.
\label{cb}
\end{eqnarray}
Clearly, $SU(2)_L\times U(1)_Y$ is spontaneously broken down to 
$U(1)_{em}$, but there is {\it no} condensation of $W$ bosons. This 
means that for such values of the lepton number it is energetically 
preferable  for the system to cancel the electric charge by means 
of the asymmetries in the quarks and charged Higgs boson, but no 
spontaneous breaking of electromagnetism takes place. 

Finally, let us consider the case of large lepton asymmetry, $n_L > 
(n_L)_2$. Now, on top of the Higgs mechanism, we have also the $W$ 
condensation \cite{l79}: 
$C\neq 0$. Notice that 
$(n_L)_2$ depends very mildly on the Higgs mass in the physically 
interesting range between $80$ GeV and $500$ GeV: $(n_L)_2\approx 
(2.0-2.5)\: T^3$. This is clearly much below the upper limit $70\: T^3$. 
Strictly speaking, for $n_L > (n_L)_1$ we have $\mu_L > T$ so that our 
analytic formulae are not exact. Thus, we have also performed 
numerical computations for the case of large chemical potentials 
and finite $\lambda$, 
which prevents exact analytic results. This amounts to including 
the terms in the effective potential of the order of $\mu^4$. Our 
findings from this standard procedure are shown in the table below, 
where we give the corrections to the critical densities calculated 
analytically. 

\begin{center}
\begin{tabular}{|c|c|c|c|}
\hline
$m_H (\hbox{\rm GeV})$ & $(n_L)_1^{num}/(n_L)_1$ & 
$(n_L)_2^{num}/(n_L)_2$ & $v_2^{num}/v_2$ \\ \hline
100 & 1.05 & 1.28 & 1.01 \\ \hline
200 & 1.07 & 1.34 & 0.92 \\ \hline
400 & 1.13 & 1.52 & 0.96 \\ \hline
600 & 1.21 & 1.66 & 1.06 \\ \hline
\end{tabular}
\end{center}
Table 1: 
The ratios between the exact numerical solutions (numerators) 
and the approximate analytic solutions (denominators) described 
in the text as functions of the Higgs mass. $v_2^{(num)}$ is the 
Higgs vev for $n_L=(n_L)_2^{(num)}$.

\vskip 0.5cm

Clearly, the numerical study confirms our analytical 
findings of symmetry breaking for large densities. Although 
the precise value of the second critical density (the first 
critical density is almost unchanged) is increased about 
$30\%$ for a reasonable values of the Higgs mass, this 
does not affect the possibility of symmetry restoration. 
Namely, the critical density remains still an order of 
magnitude below the allowed value of $70T^3$.

\vspace{0.2cm}
{\it C. The consequences: monopole and domain wall problems.} \hspace{0.5cm}
We have seen that a large enough lepton density implies symmetry breaking 
at high temperature, which opens the door for the solution of the 
monopole problem. The simplest possibility is to follow the 
scenario \cite{lp80} for the high T breaking of $U(1)_{em}$. 
The essential point here is that if the $U(1)_{em}$ symmetry is 
broken due to a large external charge, it would be broken for the whole 
parameter space of the theory and for all temperatures above $M_W$ 
all the way to the GUT scale. Thus, monopoles may never be created 
and there would obviously be no problem at all. Even if they did 
get produced they would surely have time to annihilate. 
In this sense it is only our scenario that guarantees the 
solution to the monopole problem. 
Of course one must make sure that $U(1)_{em}$ is really broken. 
If we restrict ourselves to the SM and work in the regime of 
$W$-condensation, it is not clear to us 
what the precise situation is. First of all, the 
fact that the $W$ has condensed implies the breakdown of the rotational 
invariance and the description of the formation, if any, of the monopoles 
at the  GUT scale might be completely  different from the usual one. 
Secondly, if monopoles do get 
formed, they might not annihilate rapidly enough or might not 
annihilate at all due to the anti-screening 
effects of the $W$-background. These issues are extremely 
important and certainly deserve a separate investigation. 
We would like to point out, however, that the situation is more 
transparent if we consider a simple extension of the SM where an 
electrically charged field $S$  is present (a similar extension 
would be to add another doublet). In a grand unified theory this 
singlet would be embedded in a larger representation, such as 
an $SU(5)$ \underline{10}  (a doublet would belong to 
another {\underline 5}). The idea of one singlet in addition to the 
SM Higgs was already pursued in \cite{ds92}, \cite{fkwy92}. 
However, as we axplained in the introduction, without the external 
charge this mechanism may not work \cite{gkt92}, \cite{hkr92}. 

We have explicitly checked that, for large enough lepton number, the SM 
gauge group is broken at high $T$. Moreover, since the field $S$ gets a 
VEV, $U(1)_{em}$ is spontaneously broken. More important, similarly to 
what we have described before for the SM, there exists a range of values 
of the lepton number for which the $W$-condensation does {\it not} take 
place. Under these conditions, the  monopole problem is solved. Namely,
at $T\simeq M_X$ when the GUT symmetry (say $SU(5)$) breaks down,
$U(1)_{em}$ is broken and there will be no creation of monopoles. 
It is intriguing that a realistic realization of this idea may take place 
within the Minimal Supersymmetric Standard Model (MSSM) where charged Higgs 
fields are present. The only price to pay is to accept the idea that the 
lepton number may be large enough. Once this step is made, the monopole 
problem is no longer with us. This is a remarkable result. 


What about the domain wall problem? Clearly, the presence of large 
lepton number asymmetry through the nonrestoration at high temperature 
solves the domain wall problem in an analogous manner. For example, 
this would solve the well-known domain wall problems associated with 
the spontaneous violation of CP \cite{l73} or the $Z_2$ natural 
flavour conservation symmetry \cite{gw77}.



\vspace{0.2cm}
{\it D. Summary and Outlook} \hspace{0.5cm} We have argued that 
a large lepton asymmetry in the Universe may mean an automatic 
solution of the monopole and domain wall problems through symmetry 
nonrestoration at high T. As far as the monopole problem is concerned, 
this idea works for simple extensions of the SM and in particular 
in the MSSM.

For all we know the lepton number of the Universe may be 
comparable, if not bigger than, the entropy of the Universe. The fact 
that the large lepton number can be consistent with the small baryon 
number in the context of grand unification has been pointed out 
a long time ago \cite{hk81} and recently a model for producing large 
$L$ and small $B$ has been presented \cite{ccg97}. We stress, however, 
that our findings should remain valid if, instead of the lepton 
number we consider any other conserved charge in the system under 
consideration.

\vskip 0.3cm

We would like to thank G. Dvali, M. Shaposhnikov and G.G. Ross for 
interesting discussions. This work was partially supported by the 
British Royal Society and by the Ministry of Science and Technology 
of Slovenia (B.B.) and by EEC under the TMR contract ERBFMRX-CT960090 
(G.S.). This work was completed during the Extended Workshop on 
the Highlights in Astroparticle Physics, held at ICTP from October 15 
to December 15, 1997. B.B. and A.R. thank ICTP for hospitality during 
the course of this work.

\end{document}